\documentclass[conference]{IEEEtran}
\IEEEoverridecommandlockouts
\usepackage{cite}
\usepackage{amsmath,amssymb,amsfonts}
\usepackage{algorithmic}
\usepackage{graphicx}
\usepackage{textcomp}
\usepackage{xcolor}
\usepackage{epsfig}
\usepackage{subcaption}
\usepackage{calc}
\usepackage{amstext}
\usepackage{amsthm}
\usepackage{multicol}
\usepackage{listings}
\usepackage{color}
\usepackage{cuted}
\usepackage{courier}
\usepackage{array}
\usepackage{hyperref}

\definecolor{pblue}{rgb}{0.13,0.13,1}
\definecolor{pgreen}{rgb}{0,0.5,0}
\definecolor{pred}{rgb}{0.9,0,0}
\definecolor{pgrey}{rgb}{0.46,0.45,0.48}

\def\BibTeX{{\rm B\kern-.05em{\sc i\kern-.025em b}\kern-.08em
    T\kern-.1667em\lower.7ex\hbox{E}\kern-.125emX}}
\begin{document}

\title{A Web Service Composition Method \\Based on OpenAPI Semantic Annotations}

\author{\IEEEauthorblockN{Andrei Netedu, Sabin C.~Buraga, Paul Diac, Liana \c Tuc\u ar}
\IEEEauthorblockA{\textit{Faculty of Computer Science} \\
\textit{Alexandru Ioan Cuza University of Ia\c si, Romania}\\
\{mircea.netedu, busaco, paul.diac, stefania.tucar\}@info.uaic.ro}
}

\maketitle

\begin{abstract}
Automatic Web service composition is a research direction aimed to improve the process of aggregating multiple Web services to create some new, specific functionality. The use of semantics is required as the proper semantic model with annotation standards is enabling the automation of reasoning required to solve non-trivial cases. Most previous models are limited in describing service parameters as concepts of a simple hierarchy.

Our proposed method is increasing the expressiveness at the parameter level, using concept properties that define attributes expressed by name and type. Concept properties are inherited. The paper also describes how parameters are matched to create, in an automatic manner, valid compositions. Additionally, the composition algorithm is practically used on descriptions of Web services implemented by REST APIs expressed by OpenAPI specifications. Our proposal uses knowledge models (ontologies) to enhance these OpenAPI constructs with JSON-LD semantic annotations in order to obtain better compositions for involved services. We also propose an adjusted composition algorithm that extends the semantic knowledge defined by our model.
\end{abstract}



\section{\uppercase{Introduction}}\label{sec:intro}

In the current software landscape, \emph{Web services} are the core elements of Service-Oriented Architectures (SOAs)\cite{erl2007soa} that has been already used for many years with vast popularity.
A Web Service provides a straightforward functionality defined through a public interface. In the enterprise context, this interface was traditionally  expressed in WSDL (Web Service Description Language) -- a standardized XML (Extensible Markup Language) dialect. Nowadays, Web services are usually built according to the REST (REpresentational State Transfer) architectural style~\cite{thomas2000fielding}. There are several pragmatic solutions able to describe their APIs (Application Programming Interfaces) by using lightweight formats such as JSON (JavaScript Object Notation), a well-known  data-interchange format based on ECMAScript programming language. 
In order to compose REST-based Web services, the paper proposes a novel method able to select suitable services automatically by using a knowledge-based approach described in Section~\ref{sec:formal}. Without actually executing services, valid compositions are automatically generated depending on concepts from an ontology that denotes the semantics of input parameters, properties, and output (the result). The proposed algorithm -- producing certain encouraging results -- is presented, evaluated, and discussed in Section~\ref{sec:algo}. 

We think that our proposal improves productivity and reduces costs within a complex (micro)service-based system. Additionally, the involved knowledge could be easily described, in a pragmatic way, by JSON-LD semantic constructs augmenting the OpenAPI description of each involved service, in order to offer the proper support for intelligent business operations -- see Section~\ref{sec:openapi-jsonld}. 

For a motivating case study, several experiments were also conducted by using the proposed algorithm. From a pragmatic point of view, the popular \emph{schema.org} model was chosen to convey Web services conceptual descriptions augmenting input parameters and expected results. Our approach added a suitable conceptualization that was necessary to discover compositions on cases where it could not be possible before, such as in the study presented in Section~\ref{sec:casestudy}.

The paper enumerates various related approaches (Section~\ref{sec:related}), and ends with conclusions and further directions of research.

\section{\uppercase{Problem Definition}}\label{sec:formal}

\subsection{Preliminaries}

A \emph{Web service} represents a software system designed to support inter-operable machine-to-machine interaction over a network and can be viewed as an abstract resource that represents a capability of performing tasks that form \emph{a coherent functionality} from the point of view of provider's entities and requester's entities.\footnote{Web Services Glossary, W3C Working Group Note, 2004 -- \url{https://www.w3.org/TR/ws-gloss/}}

From a computational point of view, a Web service is a set of related methods described by the same interface. This interface could be declared by adopting various specifications: 

\begin{itemize}
    \item WSDL 2.0 -- a classical Web standard\footnote{Web Services Description Language (WSDL) Version 2.0, W3C Recommendation, 2007 -- \url{https://www.w3.org/TR/wsdl20/}}  based on the XML meta-language,  
    \item OpenAPI 3.0 Specification -- a modern solution declaring the public interface of a Web service (REST API) in different formats like YAML (Yet Another Markup Language) or JSON (JavaScript Object Notation).
\end{itemize}
    
In this paper, we refer to a Web service as a \emph{single method} or the minimal endpoint that can be accessed or invoked at a time using some values as parameters -- i.e. having some prior knowledge. In our context, we also work only with ``information providing'' Web services, or \emph{stateless services} that do not alter their state and are not sensitive to the outside world states, time or any external factors. 

The main focus of the proposed method is on improving the semantic description of services motivated by the lack of means of expressing several composition techniques on previously known models and software solutions mentioned in  Section~\ref{sec:related}.

We enhanced the service composition problem by modeling semantics in the manner described below. It was inspired by our previous experience and the shortcomings we found on expressing certain natural cases of composition. More precisely, it was not possible to describe service parameters with properties or any relations/interaction between these parameters. Adding the new elements to the problem definition is also done with inspiration from the data model used by popular ontologies such as \emph{schema.org}~\cite{guha2016schema}. 

Using this approach, we managed to fix the issues that appeared in examples where the previous model failed. Our main addition is consisting of concept properties and how they are used for composition, allowing interaction between concepts and their properties based on simple constructions that increase the expressiveness.

\subsection{Proposed Formal Model}

Service parameters are defined over a set of \emph{concepts}. Let $\mathbb{C}$ be the set of all concepts that appear in a repository of services, that are all possible concepts or the problem universe.
As in the previous modeling, the concepts are first organized by using the \textit{isA} or \textit{subsumes} relation. This is a binary relation between concepts and can be considered somewhat similar to the inheritance in object-oriented programming. If a concept $c_{a}$ \textit{isA} $c_{b}$ then $c_{a}$ can substitute when there is need of a $c_{b}$. Also, for any concept $c$ in $\mathbb{C}$, there can be only one direct, more generic concept then $c$, i.e. we do not allow multiple inheritance. Obviously, \textit{isA} is transitive and, for convenience, reflexive: $c_{a}$ \textit{isA} $c_{a}$. This implies that $\mathbb{C}$ together with the \textit{isA} relation form a tree (a taxonomy) or, more generally, a forest (a set of taxonomies or a complex ontology).

\smallskip
However, the new element added to the problem are concept \emph{properties}. 
Any concept has a set of properties, possibly empty. 

Each property $p$ is a pair $\langle name, type\rangle$. The name is just an identifier of the property, and for simplicity, it can be seen as a string -- for concrete cases, this identifier is actually an IRI (Internationalized Resource Identifiers)\footnote{IRI (Internationalized Resource Identifiers -- \url{https://tools.ietf.org/html/rfc3987}}, a superset of URIs (Uniform Resource Identifiers). The type is also a concept from the same set of concepts $\mathbb{C}$ and can be considered as the range of a property.

From an ontological point of view~\cite{allemang2011semantic}, a property $p$ is defined as a relation between Web resources. The values of a property are instances of one or more concepts (classes) -- expressed by $range$ property. Any resource that has a given property is an instance of one or more concepts -- this is denoted by $domain$ property.

Properties are inherited: if $c_{a}$ \textit{isA} $c_{b}$, and $c_{b}$ has some property $\langle name_{x}, type_{x} \rangle$ then $c_{a}$ also has property $\langle name_{x}, type_{x}\rangle$. For example, if an \textit{apple} is a \textit{fruit}, and \textit{fruit} has property $\langle hasColor, Color\rangle$ expressing that \textit{fruit} instances have a color, then \textit{apple}s also must have a color.

It is important that property names do not repeat for any unrelated concepts. For any concepts that are in a \textit{isA} relation, all properties are passed to the specialization one by inheritance. This restriction is only imposed to avoid confusion and does not reduce the expressiveness, as properties can be renamed.

Syntactically, to define a property, the following are to be known: its name, its type, and the most general concept that the property can describe. For example, consider that the \textit{hasColor} property can describe the \textit{apple} concept, but also the more general \textit{fruit} concept. If concept \textit{fruit} \textit{isA} \textit{physicalObject}, the next more general concept than \textit{fruit} -- i.e., its parent in the concepts tree --, under the assumption that not all physical objects are colored, then we can say that \textit{hasColor} can most generally describe \textit{fruit}, but not any \textit{physicalObject} or any other more general concept. However, it can describe other concepts in the tree, together with all their descendants.
For simplicity, we consider further that all the properties are defined within $\mathbb{C}$, thus the concepts, \textit{isA} relation and properties structure are in $\mathbb{C}$ -- the particular ontological model.

A \textit{partially defined concept} denotes a pair \textit{(c, propSet)}, where $c$ is a concept from $\mathbb{C}$ and \textit{propSet} is a subset of the properties that $c$ has defined directly or through inheritance from more generic concepts. At some moment of time (or in some stage of a workflow), a partially defined concept describes what is currently known about a concept. It does not refer to a specific concept instance, but rather generally to the information that could potentially be found for any instance of that concept.

A Web Service \textbf{w} is defined by a pair of input and output parameters: (\textbf{w}$_{in}$, \textbf{w}$_{out}$). Both are sets of partially defined concepts. All are defined over the same structure $\mathbb{C}$, so all service providers must adhere to $\mathbb{C}$, thus adding the requirement that $\mathbb{C}$ is publicly available and defined ahead of time. In order to be able to validly call a service, all input parameters in \textbf{w}$_{in}$ must be known together with their specified required properties. After calling the service, all output parameters \textbf{w}$_{out}$ will be learned with the properties specified at output.

\smallskip
\textbf{Parameter matching}. Let \textbf{P} be a set of partially defined concepts, and \textbf{w} = (\textbf{w}$_{in}$, \textbf{w}$_{out}$) a Web service. The set \textbf{P} matches service \textbf{w} (or, equivalently, \textbf{w} is \textit{callable} if \textbf{P} is known) if and only if
$\forall$ partially defined concept \textbf{pdc} = \textit{(c, propSet)} $\in$ \textbf{w}$_{in}$,  $\exists$ \textbf{p} = \textit{($c_{spec}$, propSuperSet)} $\in$ \textbf{P} such that $c_{spec}$ \textit{isA} $c$ and $propSet \subseteq propSuperSet$. 

We further define the addition of \textbf{w}$_{out}$ to \textbf{P} as:
\newpage
\begin{strip}
  \begin{align}
     \begin{matrix}
     \textbf{P} \oplus  \textbf{w} \\ \vspace{1px} ( \hspace{1px} or \hspace{3px} \textbf{P}  \cup  \textbf{w}_{out} )      
     \end{matrix} \hspace{3px} = \hspace{3px} 
     \begin{Bmatrix}
     \Bigg(c,\bigg\{p \Big| \exists(c', propSet')\in  w_{out}  \hspace{3px} and \hspace{3px} \begin{matrix}
     c\ has \ p\\
     p \in propSet'\\
     c'\ isA\ c\\ 
     \end{matrix} \bigg\}\Bigg) \hspace{3px} \bigg| \hspace{3px} \nexists (c,propSet)\in P \\
     \end{Bmatrix} \hspace{3px} \bigcup \\
      \bigcup \hspace{3px} \begin{Bmatrix}
      \Bigg(c,propSet  \cup  \bigg\{p \Big| \exists(c',propSet')\in  w_{out} \hspace{3px} and \hspace{3px} \begin{matrix}
      c\ has \ p\\
      p \in propSet'\\ 
      c'\ isA\ c\\ 
      \end{matrix} \bigg\}\Bigg) \hspace{3px} \bigg| \hspace{3px} \exists  (c,propSet)\in P \\
      \end{Bmatrix}
  \end{align}
\end{strip}

\noindent
or the union of \textbf{w}$_{out}$ with \textbf{P} under the constraint of \textbf{P} matching \textbf{w} (defined as parameter matching above). Also, by $c$ $has$ $p$ we refer to the fact that property $p$ is stated for $c$ directly or by inheritance. \textbf{P} $\oplus$ \textbf{w} contains (1) new concepts that are in \textbf{w}$_{out}$ and (2) concepts already in \textbf{P} possibly with new properties from \textbf{w}$_{out}$ specified for corresponding concepts or their specializations. 

In words, after a call to a service, all its output parameters are selected, and for each concept together with its selected properties \textit{(c, propSet)} in \textbf{w}$_{out}$, \textit{propSet} is added to \textit{c}, \textit{c}$'$s parent in the concepts tree or the ascendants until we reach the first node that gains no new information or the root. More precisely, for each $p$ in $propSet$ we add $p$ to our knowledge base for $c$, for the parent of $c$, and so on until $p$ is no longer defined for the node we reached. The node where this process stops can differ from one $p$ property to another $p'$ property, but once the process stops for all properties in $propSet$ there is no need to go further.

\smallskip
\textbf{Chained matching}. Let \textbf{P} be a set of partially defined concepts and $(w_{1}, w_{2}, \dots, w_{k})$ an ordered list of services. We say that $\textbf{P} \oplus w_{1} \oplus w_{2} \oplus \dots \oplus w_{k}$ is a chain of matching services iff $w_{i}$ matches $\textbf{P} \oplus w_{1} \oplus w_{2} \oplus \dots \oplus w_{i-1}; \forall i = 1 \dots k$. This is the rather primitive model for multiple service calls, that is a requirement for defining the composition. For simplicity, we avoid for now more complex workflows that could handle parallel and sequential service execution constructs.

\smallskip
\textbf{Web Service Composition problem}. Given an ontology having a set of concepts $\mathbb{C}$ and a repository of Web services $W = (w_{1}, w_{2}, \dots, w_{n})$, and two sets of partially defined concepts \textbf{Init} and \textbf{Goal}, all defined over $\mathbb{C}$, find a chain of matching services $(w_{c1}, w_{c2}, \dots w_{ck})$ such that $(\emptyset, $\textbf{Init}$) \oplus w_{c1} \oplus w_{c2} \oplus \dots \oplus w_{ck} \oplus ($\textbf{Goal}$, \emptyset)$.

The $(\emptyset, \textbf{Init})$ and $(\textbf{Goal},$ $\emptyset)$ are just short ways of writing the initially known and finally required parameters, by using mock services. 

We can also imagine (\textbf{Init}, \textbf{Goal}) as a Web (micro-)service -- in this context, the problem requires finding an ``implementation'' of a Web (micro-)service using the services available in a certain development environment (e.g., a public or private repository).

\section{\uppercase{Automatic Service Composition Algorithm}}\label{sec:algo}

\subsection{Algorithm Description}

The proposed algorithm is intended to describe a generic solution that generates a valid composition. Tough it considers some basic optimizations like special data structures and indexes, there are many ways in which it can be improved, so we shortly describe some of them after the basic algorithm description.

In a simplified form, considered main entities have the following structure:

\begin{lstlisting}[basicstyle=\footnotesize\ttfamily]
class Concept {   // full or partial type
  Uri uri;        // unique identifier
  String name;    // a label
  Concept parent; // isA relation
  // proper and inherited properties
  Set<Property> properties; 
}

class Property {
  Uri uri;        // unique identifier
  String name;    // a label
  Concept type;   // property's range
}

class WebService {
  Uri uri;
  String name;
  // I/O parameters
  Set<Concept> in, out;
}	
\end{lstlisting}

\smallskip
Global data structures that are most important and often used by the algorithm are presented below:

\begin{lstlisting}[basicstyle=\small\ttfamily]
// knowledge: concepts, isA, properties
Set<Concept> C;

// repository of services
Set <WebService> webServices;
// user's query as two fictive services
WebService Init, Goal;

// the known partially defined concepts
// known.get(c) = concept's known properties
Map <Concept,Set<Property>> known;

// required.get(c).get(p) services that
// have property p of concept C in input
Map <Concept,Map<Property, Set<WebService>>> required; 


// remaining.get(w).get(c) = 
// properties of concept C 
// necessary to call the W service
Map <WebService, Map<Concept, Set<Property>>> remaining;

// services for which all input is already known
Set<WebService> callableServices; 
\end{lstlisting}

The algorithm described next uses the above structures, and is composed of three methods: initialization, the main composition search method which calls the last (utility) method, that updates the knowledge base with the new outputs learned form a service call. 

Several obvious instructions are skipped for simplicity like parsing input data and initializing empty containers.

\begin{lstlisting}[basicstyle=\small,escapeinside={[}{]}\ttfamily]
void initialize() {
 // read C, services, query as (Init.out, Goal.in)
 Init.in = Goal.out = [ $\emptyset$ ];
 webServices.add(Goal);
 
 for (WebService ws : webServices) {
  for (Concept c : ws.in) {
   for (Property p : c.properties) {    
    required.get(c).get(p).add(ws);
    remaining.get(ws).get(c).add(p);       
   } // container creation skipped   
[ \vspace{-15px}]
  }
[ \vspace{-15px}]
 }
[ \vspace{-15px}]
}
\end{lstlisting}

After reading the problem instance, the described data structures have to be \emph{loaded}. \textbf{Init} and \textbf{Goal} can be used as web services to reduce the implementation size, if they are initialized as above. Then, for each parameter in service's input, we add the corresponding concepts with their specified properties to the \emph{indexes} (maps) that efficiently get the services that have those properties at input and the properties that remain yet unknown but required to validly call a service.

\begin{lstlisting}[basicstyle=\small,escapeinside={[}{]}\ttfamily]
List<WebService> findComp(WebService Init, Goal) {
 List<WebService> composition; //result
 callService(Init); // learn initial
 
 while (!(required.get(Goal).isEmpty() ||    
          callableServices.isEmpty())) {
  WebService ws = callableServices.first();
  callableServices.remove(ws);
  composition.add(ws);
  callWebService(ws);
 }
 if (!remaining.get(Goal).isEmpty()) {
   return null; // no valid composition
 } else { 
   return composition; 
 }
[ \vspace{-15px}]
}
\end{lstlisting}

The main method that searches for a valid composition satisfying user's query is \emph{findComp()}. The result is simplified for now as an ordered list of services. As long as the \textbf{Goal} service is not yet callable, but we can call any other new service, we pick the \emph{first} service from the \emph{callableServices} set. Then we simulate its call, basically by learning all its output parameter information, with the help of \emph{callWebService()} method below. We add the selected service to the composition and remove it from \emph{callableServices} so it won't get called again. If \emph{callableServices} empties before reaching the \textbf{Goal}, then the query is unsolvable.

\begin{lstlisting}[basicstyle=\small,escapeinside={[}{]}\ttfamily]
// 'discover' and expand the output of a Web service
void callWebService(WebService ws) {
 for (Concept c : ws.out) {
  Concept cp = c; // concept goes up in tree
  // true if anything new was learned at level
  boolean added = true;
  
  while (added && cp != null) {
   added = false;
   for (Property p : c.properties) {
    if (cp.properties.contains(p) &&
        !known.get(cp).contains(p)) {
     added = true;
     // learn p at cp level:
     known.get(cp).add(p);
     for (WebService ws: required.get(cp).get(p)) {
      remaining.get(ws).get(cp).remove(p);
      if (remaining.get(ws).get(cp) .isEmpty()) {
       // all properties of cp in ws.in are known
       remaining.get(ws).remove(cp);
      }
      if (remaining.get(ws). isEmpty()) {
       // all concepts in ws.in known
       callableServices.add(ws);       
      }
[ \vspace{-15px}]
     }
[ \vspace{-15px}]
    }
[ \vspace{-15px}]
   }  
   cp = cp.parent;
  }  
[ \vspace{-15px}]
 }
[ \vspace{-15px}]
}
\end{lstlisting}

When calling a Web service, its output is learned and also expanded, i.e. we mark as learned properties also for more generic concepts. This improves the algorithm's complexity, as it is better to prepare detection of newly callable services than to search for them by iteration after any call. This is possible by marking in the tree the known properties for each level and iterating only to levels that get any new information, as higher current service's output would be already known. 

The optimization also comes from the fact that all services with inputs with these properties are hence updated only once (for each learned concept and property). As it gets learned, information is removed from the \emph{remaining} data structure first at the property level and then at the concept level. When there's no property of any concept left to learn, the service gets callable. This can happen only once per service. The main loop might stop before reaching the tree root if at some generalization level, all current services' output was already known, and this is determined by the \emph{added} flag variable.

\subsection{Possible Improvements}

There are several improvements that can be added to the algorithm. One important metric that the algorithm does not consider is the size of the produced composition. As can be seen from the overview description above, the solution is both deterministic and of polynomial time complexity. This is possible because the length of the composition is not a necessary minimum. Finding the shortest composition is \texttt{NP-Hard} even for problem definitions that do not model semantics. The proposed model introduces \textit{properties}, but this addition does not significantly increase the computational problem complexity. Nevertheless, even if the shortest composition is hard to find, there are at least two simple ways to favor finding shorter compositions with good results.

One is based on the observation that when a service is called, it is chosen from possibly multiple callable services. This choice is not guided by any criteria. It is possible to add at least a simple \textit{heuristic score} to each service that would estimate how useful is the information gained by that service call. To take this even further, this score can be updated for the remaining services when information is learned.

Another improvement is based on the observation that services can be added to the composition even if they might produce no useful information -- there is no condition check that they add anything new, or the information produced could also be found from services added later to the composition. To mitigate this, another algorithm can be implemented that would search the resulting composition \textit{backward} and remove services that proved useless in the end.

Both of the above improvements can have an impact on the running time as well, as the algorithm stops when the goal is reached.

\subsection{Empiric Evaluation}

To test the algorithm implementation, we first used some handwritten examples, including the one in Section \ref{sec:casestudy}. Then, we implemented a random test generator to analyze the performance on larger instances.

The generator first creates a conceptual model (a dummy ontology) with random concepts and properties and then a service repository based on the generated ontology. Service parameters are chosen from direct and inherited properties. In the last step, an ordered list of services is altered by rebuilding each service input. The first one gets its input from \textbf{Init}, which is randomly generated at the beginning. Each other service in order gets its input rebuilt from all previous services outputs, or from valid properties of generalizations of those concepts. Finally, \textbf{Goal} has assigned a subset of the set of all outputs of the services in the list. The total number of services and the list size are input parameters for the generator. The intended dependency between services in the list is not guaranteed, so shorter compositions could potentially exist.

\setlength\extrarowheight{2pt}
\begin{table*}[ht]
  \centering
  \caption{Algorithm run times and resulting composition size on random generated instances.}
  \label{tab:runtimes}
  \begin{tabular}{|c|c|c|c|c|}
    \hline
    ontology size & repository & run time & result composition & dependency list \\
    (\#concepts + \#properties) & size: \#services & in seconds & size: \#services & size: \#services\\
    \hline
    10 (5 + 5) & 10 & 0.002 & \textbf{3} & 5 \\
    20 (10 + 10) & 20 & 0.003 & \textbf{4} & 10 \\
    50 (30 + 20) & 20 & 0.007 & \textbf{12} & 20 \\
    20 (10 + 10) & 50 & 0.011 & \textbf{6} & 20 \\
    \hline
\end{tabular}
\end{table*}

Table~\ref{tab:runtimes} shows the algorithm performance. The first two columns briefly describe the input size, by the number of concepts and properties in the generated ontology and total number of services in the repository. The column \emph{result composition size} measures the length of the composition found by the algorithm. The last column, \emph{dependency list size}, measures the length of the composition generated by the tests generator algorithm. The \emph{dependency list} constitutes a valid composition, hidden within the repository and may contain useless services as the \emph{dependency} is not guaranteed.

\section{\uppercase{Extending OpenAPI with JSON-LD constructs}}\label{sec:openapi-jsonld}

Another aim of this research is to show how current OpenAPI specification\footnote{OpenAPI Specification -- \url{https://github.com/OAI/OpenAPI-Specification}} and our proposed extension to the JSON-LD model can be used for automatic Web (micro-)service composition.

\subsection{From Formalism to Semantic Descriptions}

As a first step, we applied the above mathematical model and algorithm for a set of Web services defined with the help of  OpenAPI specification expressed by JSON (JavaScript Object Notation)\footnote{JSON (JavaScript Object Notation) -- \url{https://json.org/}} constructs.

OpenAPI specification is used to describe Web services (APIs) aligned to the REST (REpresentational State Transfer) architectural style~\cite{thomas2000fielding}. This specification defines in a standardized manner a meta-model to declare the interfaces to RESTful APIs in a programming language-agnostic manner.

These APIs correspond to a set of Web services that forms a repository. For our conducted experiments, we considered an API as a collection of services where each different URI path of the API denoted a different service -- this can also be useful in the context of microservices. Thus, we can group services based on information related to the location of the group of services. In practice, this is very useful as generally a Web server would likely host a multitude of Web services. 

We used OpenAPI specification to describe the input and output parameters of each different service. Those parameters are Web resources that represent, in our mathematical model, partially defined concepts. OpenAPI also helps us match parameters on a syntactic level by specifying the data types of the parameters.

\smallskip
\textbf{Example.} As a real-life case, the public REST API (Web service) provided by Lyft\footnote{Lyft API -- \url{https://developer.lyft.com/docs}} is described. For the operation of getting details about an authenticated user (\texttt{GET /profile}), the result -- i.e., \textit{w.out} in our mathematical model and, in fact, output data as a JSON object composed by various properties -- is the following:

\begin{itemize}
    \item \textit{id}: string -- Authenticated user's identifier.
    \item \textit{first\_name}: string -- First name.
    \item \textit{last\_name}: string -- Last name.
    \item \textit{has\_taken\_a\_ride}: boolean -- Indicates whether the user has taken at least one Lyft ride.
\end{itemize}

Each property can have as a result -- the $range$ mentioned in Section~\ref{sec:formal} -- a datatype denoted by a concept, according to the proposed mathematical model. The Web resources processed by a service are then linked to assertions embedded into JSON-LD descriptions. All JSON-LD statements could be converted into RDF (Resource Description Framework) or OWL (Web Ontology Language)~\cite{allemang2011semantic} to be further processed, including automated reasoning.

For this example, the \textit{first\_name} property of the returned JSON object could be mapped -- via JSON-LD annotations -- to the \textit{givenName} property of \textit{Person} concept (class) defined by the schema.org conceptual model or -- as an alternative -- by FOAF (Friend Of A Friend) vocabulary\footnote{FOAF (Friend Of A Friend) Vocabulary Specification -- \url{http://xmlns.com/foaf/spec/}}. Additionally, the output parameter -- i.e., the returned JSON object -- could be seen as an instance of \textit{Person} class. This approach could be further refined, if a \textit{Customer} and \textit{Driver} class are defined as a subclass of \textit{Person} concept. The \textit{subclass} relation from the (onto)logical model -- formally written as \textit{Customer} $\sqsubseteq$ \textit{Person} -- is equivalent to the \textit{isA} relation of the mathematical formalism presented in Section~\ref{sec:formal}. Using this taxonomy of classes, the algorithm could select the Web service as a candidate solution for Web service composition.

\smallskip
Secondly, we chose the JSON-LD model to describe the entities used by a group of services. For each different Web resource specified in the OpenAPI description of considered Web services, a corresponding semantic description could be stated. This semantic description attaches to each Web resource a URI specifying a concept in an ontology. Similarly, for each property of the corresponding Web resource.

\subsection{Generic Approach}

Generally, considering an OpenAPI specification stored in a JSON document:

\begin{lstlisting}[basicstyle=\small,escapeinside={<}{>}\ttfamily]
{ 
< \vspace{-12px}>
  "openapi": "3.0.0",
  "paths": {
    "/resource": {
      "get": {
        "operationId": "service",
        "parameters": [ {
          "name": "parameter",
          "in": "query",
          "schema": { "type": "string" }
        } ],
        "responses": {
          "200": {
            "description": "Success",
            "schema": {
              "$ref":"#/definitions/Response"
            }
< \vspace{-13px}>
        }
< \vspace{-13px}>
     } /* a response object composed by properties and values, e.g. { response: string } */
< \vspace{-15px}>
   }
< \vspace{-15px}>
  }
< \vspace{-15px}>
 }
< \vspace{-15px}>
}
\end{lstlisting}

\noindent 
an abstract JSON-LD annotation has the following form, where the context is a knowledge model used to denote Web service entities and the \textit{Concept1} and \textit{Concept2} classes are used for each parameter instead of a generic JSON datatype (in this case, a string).

\begin{lstlisting}[basicstyle=\small,escapeinside={<}{>}\ttfamily]
{
 "@context": "http://ontology.info/",
 "@id": "parameter",
 "@type": "Concept1" // w.in
 "response": {
  "@type": "Concept2" // w.out
 }
}
\end{lstlisting}

Furthermore, a convenient mapping -- a-priori given or automatically generated -- between JSON datatypes (string, number, boolean, object, array) and ontological concepts could be attached as meta-data for the considered set of Web services in order to facilitate the matching process. This enhancement is inspired by ontology alignment strategies~\cite{shvaiko2013ontology}. The mapping itself could be directly expressed in JSON-LD via \texttt{@context} construct used to map terms (in our case, datatype names and/or property names) to concepts denoted by URIs. The concepts, properties, restrictions, and related entities (such as individuals and annotations) form the knowledge base -- usually, specified by using OWL and RDF standards~\cite{allemang2011semantic}.

\section{\uppercase{Case Study: A Transport Agency}}\label{sec:casestudy}

To illustrate the benefits of our approach, we have considered the following example, according to the problem definition in Section~\ref{sec:formal}. Also, we have used concepts from schema.org. The services' interfaces are stored in an OpenAPI compliant document. The resources (service parameters) are described via JSON-LD by using the method exposed in Section~\ref{sec:openapi-jsonld}.

The case study specifies the car company operating processes via web services. The car company services are simplistic and perform small tasks that can significantly benefit from a web service composition solution. Describing the services using OpenAPI and JSON-LD, the approach showcases how easy it is to represent complex relations between resources by combining object-oriented concepts in both structure and semantics. On top of this, because the REST services are similar on an architectural level, the solution guarantees that the situation presented in the case study can easily be extended and applied in real-world scenarios. 
For the conducted experiment, our example contains six services, several resources, but also the query to be solved by a composition of all six services in the correct order.

The scenario is the following. We are supposing that a customer needs a \emph{vehicle} to transport a given \emph{payload}. This person knows his/her current \emph{GeoLocation(latitude, longitude)} and a time frame \emph{Action(startTime, endTime)} in which the transport should arrive. The \textbf{Goal} is to obtain the \emph{Action(location)} where the \emph{vehicle} will arrive. 
The six services -- each of them implementing a single operation -- are specified below:

\begin{lstlisting}[basicstyle=\small,escapeinside={<}{>}\ttfamily]
getCountryFromLocation
 in  = GeoLocation(lat,lon)
 out = Country(name)
 < \vspace{-16px}>
    
getTransportCompany
 in  = AdministrativeArea(name)
 out = Organization(name)
< \vspace{-13px}>
    
getClosestCity
 in  = GeoLocation(lat,lon)
 out = City(name)
< \vspace{-16px}>
    
getLocalSubsidiary
 in  = Organization(name), City(name)
 out = LocalBusiness(email)
< \vspace{-13px}>
 
getVehicle
 in = Vehicle(payload),
      LocalBusiness(email)
 out= Vehicle(
       vehicleIdentificationNumber)
< \vspace{-13px}>

makeArrangements
 in = Vehicle(
       vehicleIdentificationNumber),
      Organization(name,email), 
      Action(startTime,endTime)
 out = Action(location)
\end{lstlisting}

In OpenAPI terms, a HTTP GET method is defined to obtain a JSON representation of the desired Web resource, for each \textit{getResource} operation -- i.e. using \texttt{GET /country} with \emph{GeoLocation} as input parameter and \emph{Country} as output. Without JSON-LD constructs, these parameters have regular JSON datatypes like string or number. 

As defined by schema.org, \emph{LocalBusiness}\footnote{LocalBusiness, a particular physical business or branch of an organization -- \url{https://schema.org/LocalBusiness}} \textit{isA} \emph{Organization} -- or, equivalent, in (onto)logic terms: \emph{LocalBusiness} $\sqsubseteq$ \emph{Organization}. Similarly, \emph{Country} \textit{isA} \emph{AdministrativeArea}. 

\smallskip
A valid composition satisfying the user request can consist of the services in the following order: \textbf{Init} $\rightarrow$ \emph{getCountryFromLocation} $\rightarrow$ \emph{getTransportCompany} $\rightarrow$ \emph{getClosestCity} $\rightarrow$ \emph{getLocalSubsidiary} $\rightarrow$ \emph{getVehicle} $\rightarrow$ \emph{makeArrangements} $\rightarrow$ \textbf{Goal}. The order is relevant, but not unique in this case. This can be verified by considering all resources added by each service and also by the use of the \textit{isA} relation. For example, \emph{LocalBusiness(email)} can be used as \emph{Organization(email)}.

\smallskip
The Java implementation of the algorithm from Section~\ref{sec:algo} reads the OpenAPI specification document describing the above scenario and finds the highlighted composition based on schema.org conceptual model. We also validated the OpenAPI and JSON-LD files using available public software tools included in Swagger Editor\footnote{Swagger Editor -- \url{http://editor.swagger.io/}} and JSON-LD Playground\footnote{JSON-LD Playground -- \url{https://json-ld.org/playground/}}.

\section{\uppercase{Related Work}}\label{sec:related}

Various approaches~\cite{rao2004survey,milanovic2004current,blake2006wsc} were considered to tackle the problem of Web service composition in the case of ``classical'' specification of Web services by using the old SOAP, WSDL (Web Service Description Language), and additional WS-* standards and initiatives~\cite{erl2007soa}. Also, several semantic-based formal, conceptual, and software solutions are proposed by using DAML-S and its successor OWL-S\footnote{OWL-S: Semantic Markup for Web Services -- \url{https://www.w3.org/Submission/OWL-S/}} service ontologies (in the present, they are almost neglected by the existing communities of practice) and language extensions to Web service descriptions. 

For instance, the following initiatives, methods, and software solutions could be mentioned:

\begin{itemize}
    \item A mathematical model of the semantic Web service composition problem considering AI planning and causal link matrices~\cite{lecue2006formal}.
    \item A linear programming strategy for service selection scheme considering non-functional QoS (Quality of Service) attributes~\cite{cardellini2007flow}.
    \item Automated discovery, interaction, and composition of the Web services that are semantically described by DAML-S ontological constructs~\cite{sycara2003automated}.
    \item A high-level conceptual architecture and model for Web service deployment, providing support for  service interaction, mediation, and composition of services~\cite{preist2004conceptual}.
    \item A process ontology (called OWL-S) to describe various aspects of Web services using SOAP protocol~\cite{martin2007bringing}.
    \item A service composition planner by using hybrid artificial intelligence techniques and OWL-S model~\cite{klusch2005semantic}.
    \item A hybrid framework which achieves the semantic web service matchmaking by using fuzzy logic and OWL-S statements~\cite{fenza2008hybrid}.
    \item An automated software tool using MDA (Model-Driven Architecture) techniques to generate OWL-S descriptions from  UML models~\cite{timm2005model}.
    \item Various efforts for annotating SOAP-based Web service descriptions by using semantic approaches -- e.g., SAWSDL (Semantic Annotations for WSDL and XML Schema) -- within the METEOR-S system~\cite{sheth2008semantics}.
\end{itemize}

In contrast, there are relatively few recent proposals focused on resolving the problem of automatic service composition in the context of the new pragmatic way of describing the public REST APIs by using OpenAPI specification. Several solutions and tools are covered in~\cite{garriga2016restful} and~\cite{lemos2016web}. From the modeling service compositions perspective, our formal model presents several similarities to the solutions proposed by~\cite{baccar2018declarative} and~\cite{cremaschi2018practical}. 

Concerning enhancing Web services with semantic descriptions~\cite{verborgh2014survey}, several recent initiatives considering OpenAPI-based approaches are focused on:

\begin{itemize}
    \item Extending the OpenAPI specification with a meta-model giving developers proper abstractions to write reusable code and allowing support for design-time consistency verification~\cite{sferruzza2018extending}. This approach is not using any semantic descriptions and does not provide support for reasoning via JSON-LD annotations.
    \item Using fuzzy inference methods to match services considering QoS metrics \cite{wang2006fuzzy}, or denoted by OpenAPI specifications~\cite{peng2018fuzzy}.
    \item Generating, in an automatic manner, the suitable GraphQL-based wrappers for REST APIs described by OpenAPI documents~\cite{wittern2018generating}.
    \item Using the RESTdesc\footnote{RESTdesc -- \url{http://restdesc.org/}} for service composition and invocation processes in the context of the Internet of Things -- e.g., smart sensors~\cite{ventura2015autonomous}.
    \item Capturing the semantics and relationships of REST APIs by using a simplified description model (Linked REST APIs), to automatically discover, compose, and orchestrate Web services~\cite{serrano2018automated}.
\end{itemize}

\section{\uppercase{Conclusion and Future Work}}\label{sec:conclusions}

The paper focused on REST-based Web services composition by using a straightforward method that adopts a conceptual approach for modeling semantics of (micro)service parameters. Starting with a formal model described in Section~\ref{sec:formal}, an automatic service composition algorithm was proposed and evaluated -- see Section~\ref{sec:algo}. 

To prove the feasibility of the proposed formalism, we extended OpenAPI description of stateless services with JSON-LD constructs, in order to conceptually explain the involved entities (resources) of the service's interface. This new method was detailed in Section~\ref{sec:openapi-jsonld}. 

For practical reasons, we adopted the popular \emph{schema.org} model to quickly determine taxonomic relationships between concepts according to the described algorithm. This was exemplified by the case study presented in Section~\ref{sec:casestudy}. Our approach is general enough to choose convenient ontologies for each specific set of composable (micro-)services. We consider that our proposal is also a suitable solution for service-based business Web applications deployed in various cloud computing platforms.   

An alternative approach is to use the SHACL (Shapes Constraint Language) model\footnote{Shapes Constraint Language (SHACL), W3C Recommendation, 2017 -- \url{https://www.w3.org/TR/shacl/}} to specify certain restrictions on RDF and, equally, on JSON-LD data -- this direction of research is to be investigated in the near future in order to provide support for accessing semantically enriched digital content~\cite{levinatowards}.

We are aware that the proposed method presents several shortcomings -- e.g., lack of support regarding service meta-data such as quality of service, various restrictions, work and data-flows, and others. These aspects will be considered, formalized, and implemented in the next stages of our research.

\bibliographystyle{IEEEtran}
\bibliography{services}

\end{document}